\documentstyle[aps,prl,epsfig,preprint,upcite]{revtex}
\begin{document}
\pagestyle{myheadings}
\markboth{Helbing/Huberman: Moving Like a Solid Block}
{Helbing/Huberman: Moving Like a Solid Block}
\title{\mbox{}\\[-2.2cm]Coherent Moving States in Highway
Traffic\\[-0cm]}
\author{Dirk Helbing$^*$ and Bernardo A. Huberman$^+$}
\address{$^*$ II. Institute of Theoretical Physics, University of Stuttgart,\\
Pfaffenwaldring 57/III, 70550 Stuttgart, Germany\\
$^+$ Xerox PARC, 3333 Coyote Hill Road, Palo Alto, CA 94304, USA\\
{\tt helbing@theo2.physik.uni-stuttgart.de, huberman@parc.xerox.com}\\[8mm]}
\maketitle
\draft
{\bf Recent advances in multiagent simulations\,\cite{Nag}$^-$\cite{TGF} 
have made possible the
study of realistic traffic patterns and allow to test 
theories based on driver behaviour\,\cite{TGF}$^-$\cite{book}. 
Such simulations also display various 
empirical features of traffic flows\,\cite{May}, 
and are used to design traffic controls 
that maximise the throughput of vehicles in heavily
transited highways. In addition to its 
intrinsic economic value\,\cite{Small}, vehicular traffic 
is of interest because it may throw light on some 
social phenomena where diverse individuals competitively try to
maximise their own utilities 
under certain constraints\,\cite{SciAm,QuantSoz}. 
\par
In this paper, we present simulation results that
point to the existence of cooperative,
coherent states arising from competitive 
interactions that lead to
a new phenomenon in heterogeneous highway traffic.
As the density of vehicles 
increases, their interactions cause
a transition into a highly correlated state 
in which all vehicles practically move with the same speed, analogous
to the motion of a solid block. This state is associated with a reduced
lane changing rate and a safe, high and stable flow. It 
disappears as the vehicle density exceeds a critical value. The
effect is observed in recent evaluations of Dutch traffic data.}
\par %\clearpage
In many social situations, decisions made by individuals 
lead to externalities, which may be very regular,
even without a global coordinator. 
In traffic, these decisions concern when to 
accelerate or brake, pass, or enter into a
heavily transited multi-lane road\,\cite{Dag}$^-$\cite{zwei}, 
while trying to get ahead as fast as possible, but
safely, under the constraints imposed by physical limitations and traffic
rules. At times this behaviorial rules give rise to very regular traffic patterns as
exemplified by 
the universal characteristics of moving jam fronts\,\cite{Kerner1} 
or synchronised congested traffic\,\cite{Kerner2,Lett}. 
These phenomena are in contrast with 
usual social dilemmas, where cooperation in order to achieve
a desirable collective behaviour hinges on having
{\em small} groups or {\em long} time horizons\,\cite{Ax,SciAm}.
\par
\begin{figure}
\begin{center} \hspace*{-0.5\unitlength}
\epsfig{height=8\unitlength, angle=-90,
      file=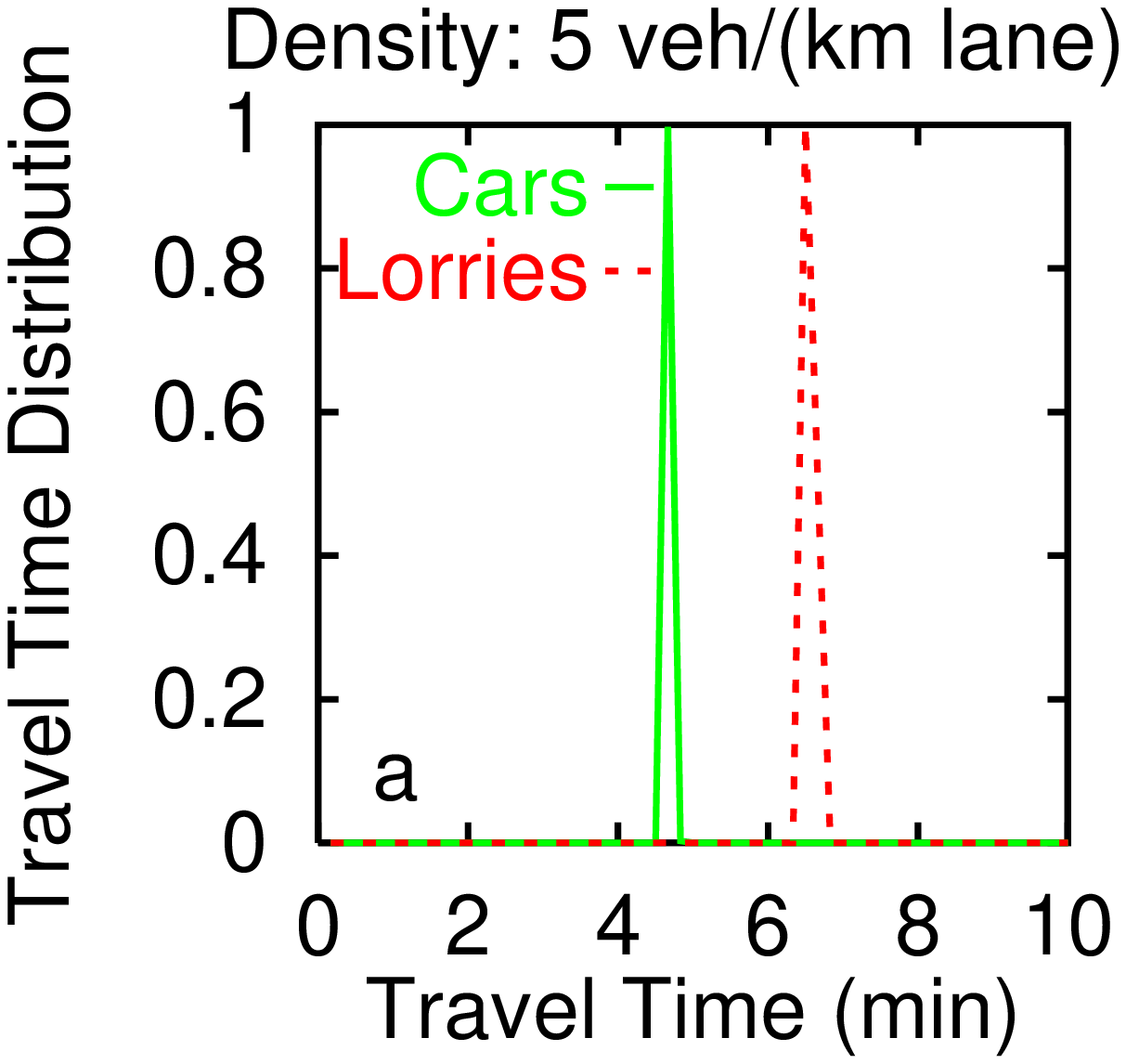} \hspace*{-0.5\unitlength}
\epsfig{height=8\unitlength, angle=-90,
      file=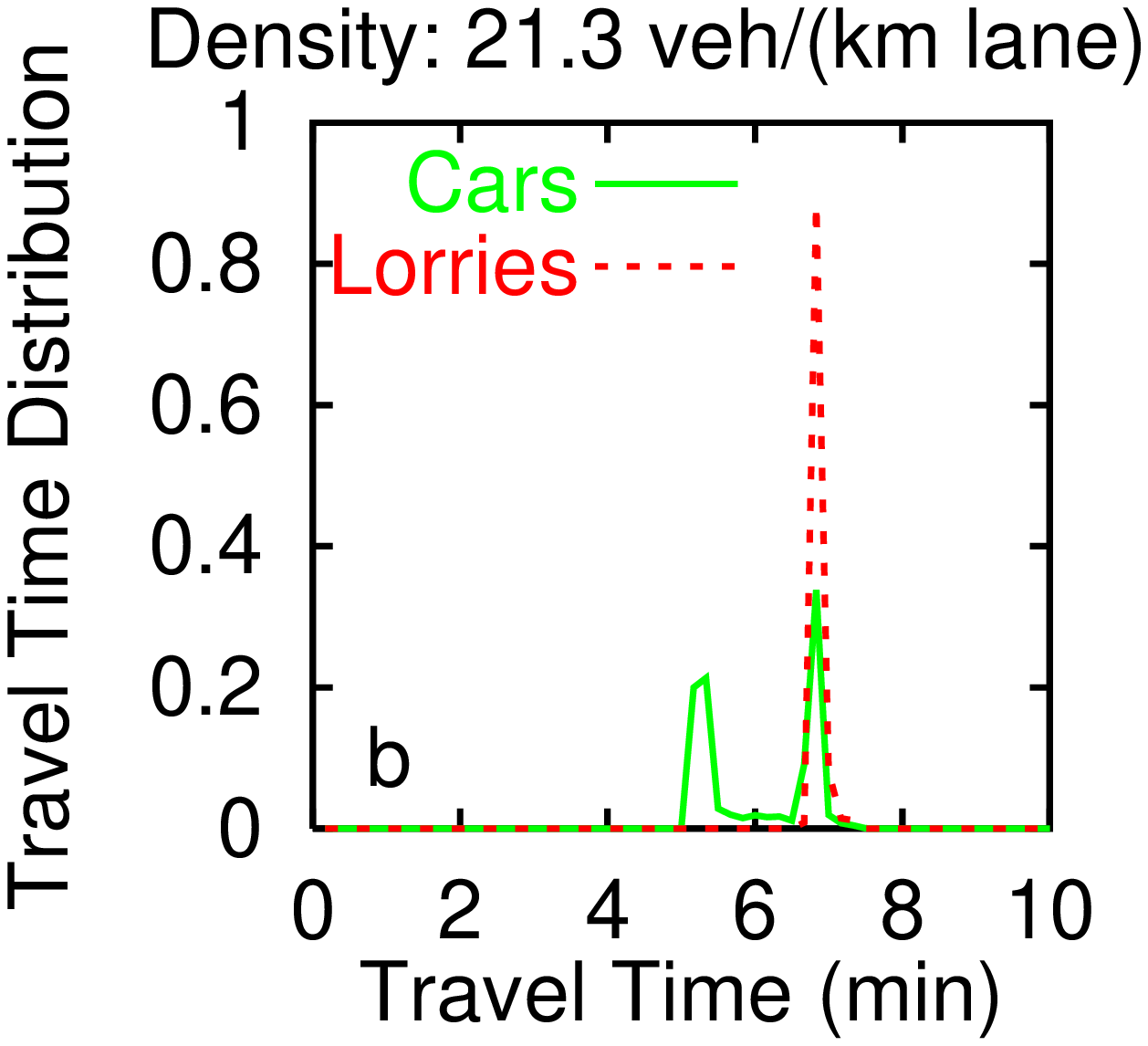} \\ \hspace*{-0.5\unitlength}
\epsfig{height=8\unitlength, angle=-90,
      file=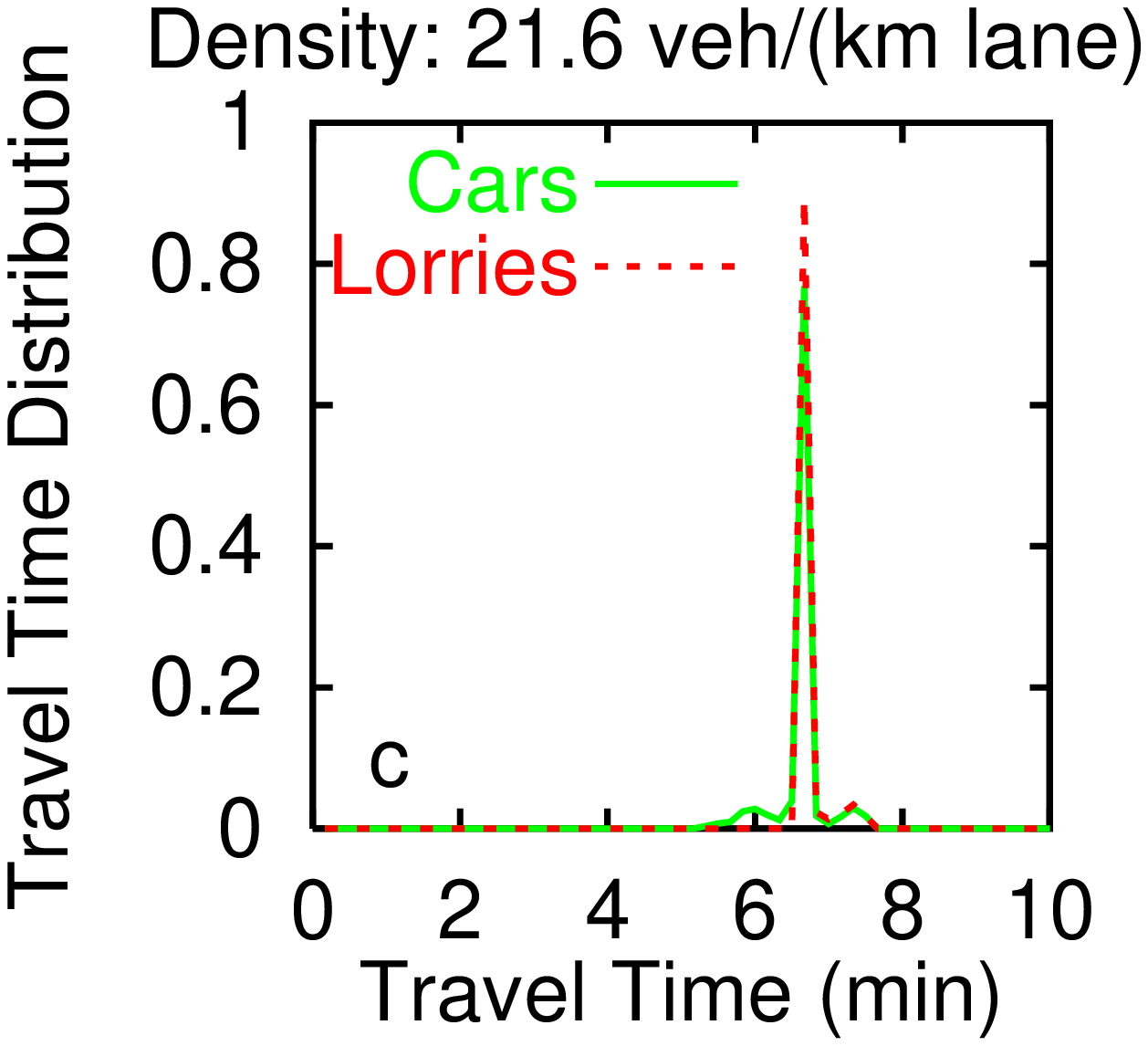} \hspace*{-0.5\unitlength}
\epsfig{height=8\unitlength, angle=-90,
      file=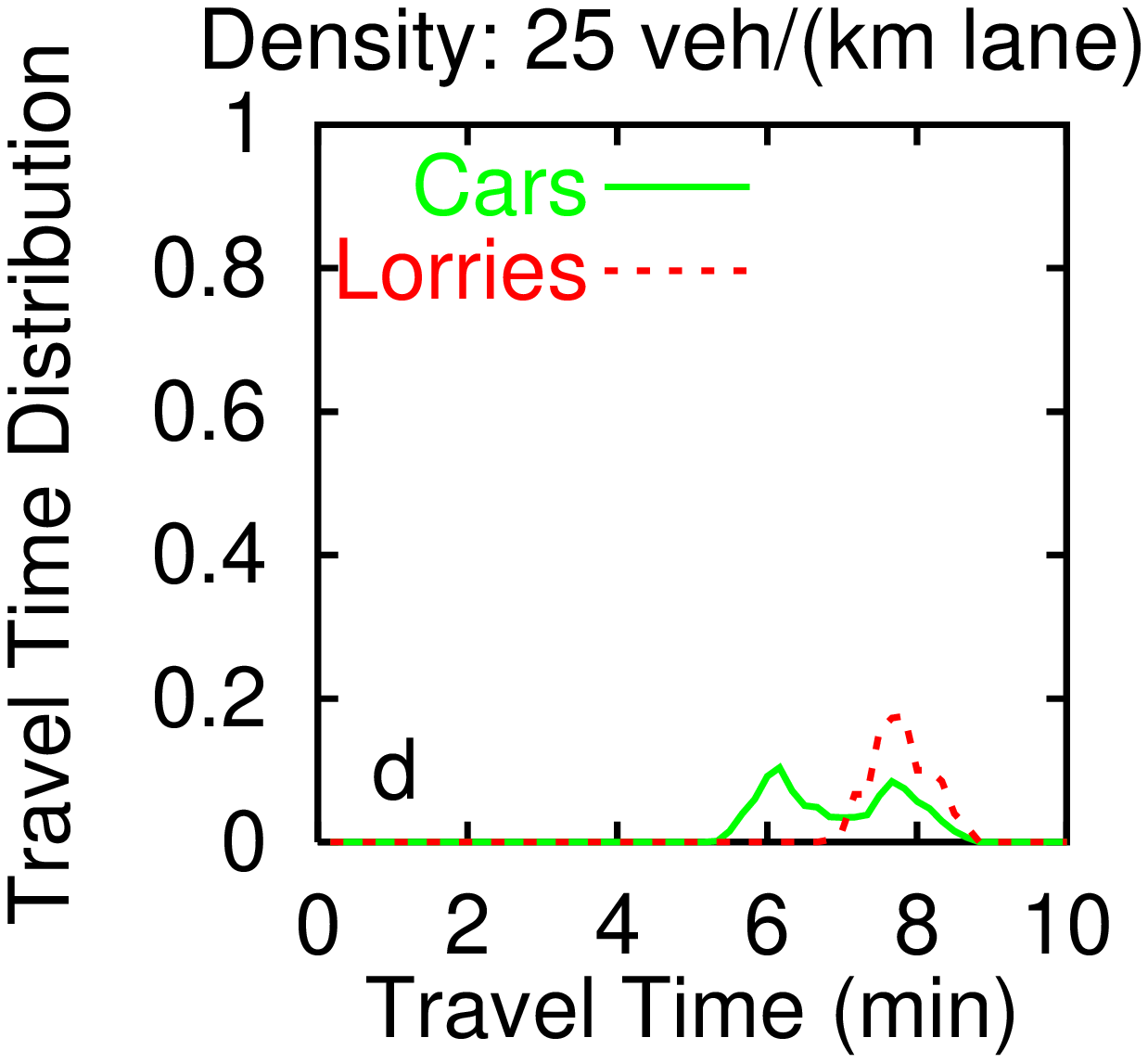}
\end{center}
\caption[]{Simulated travel time distributions 
for a circular one-way two-lane 
highway of 10 kilometer length. 
The chosen model parameters 
$\lambda = 0.77$, $p=0.05$, and $m=2$, together with the desired
velocity functions displayed in Fig.~\ref{F2}, yield a good 
representation of the dynamics on fast lanes of Dutch
freeways\,\cite{zellauto}.\\
We considered a scenario of about 2.5\% randomly selected, 
identical lorries of length 7.5\,m with a maximum velocity of about
90\,km/h moving among 97,5\% identical cars with 5\,m length and a maximum
velocity of about 125\,km/h. 
(a) At small densities, the travel time distribution has
two narrow peaks at the maximum velocities
of cars and lorries, since cars can overtake lorries
without prior slowdowns.
(b) With increasing but still moderate density,
the travel times of lorries remain unchanged, 
as their slow speed implies large average headways. 
In contrast, the average travel time 
of cars grows. As a lack of sufficiently large gaps 
may prevent immediate lane changing,
some cars will have to temporarily slow down to the lorries' speed.
The resulting higher relative velocity to the vehicles in the adjacent lane 
makes overtaking more difficult, so they may get `trapped' behind lorries 
for a long time. Therefore, the travel time
distribution develops a second peak around the lorry peak. 
(c) At a certain density, the peak of unobstructed cars
disappears, and the travel time distributions of cars and lorries become
almost identical, with a small dispersion. 
(d) If the density is further increased,
this highly correlated state of motion breaks down, and a
broad distribution of travel times results. 
\label{F1}}
\end{figure}
\begin{figure}
\begin{center}
\begin{picture}(8,3.3)
\put(0,3.3){\epsfig{height=8\unitlength, angle=-90,
      file=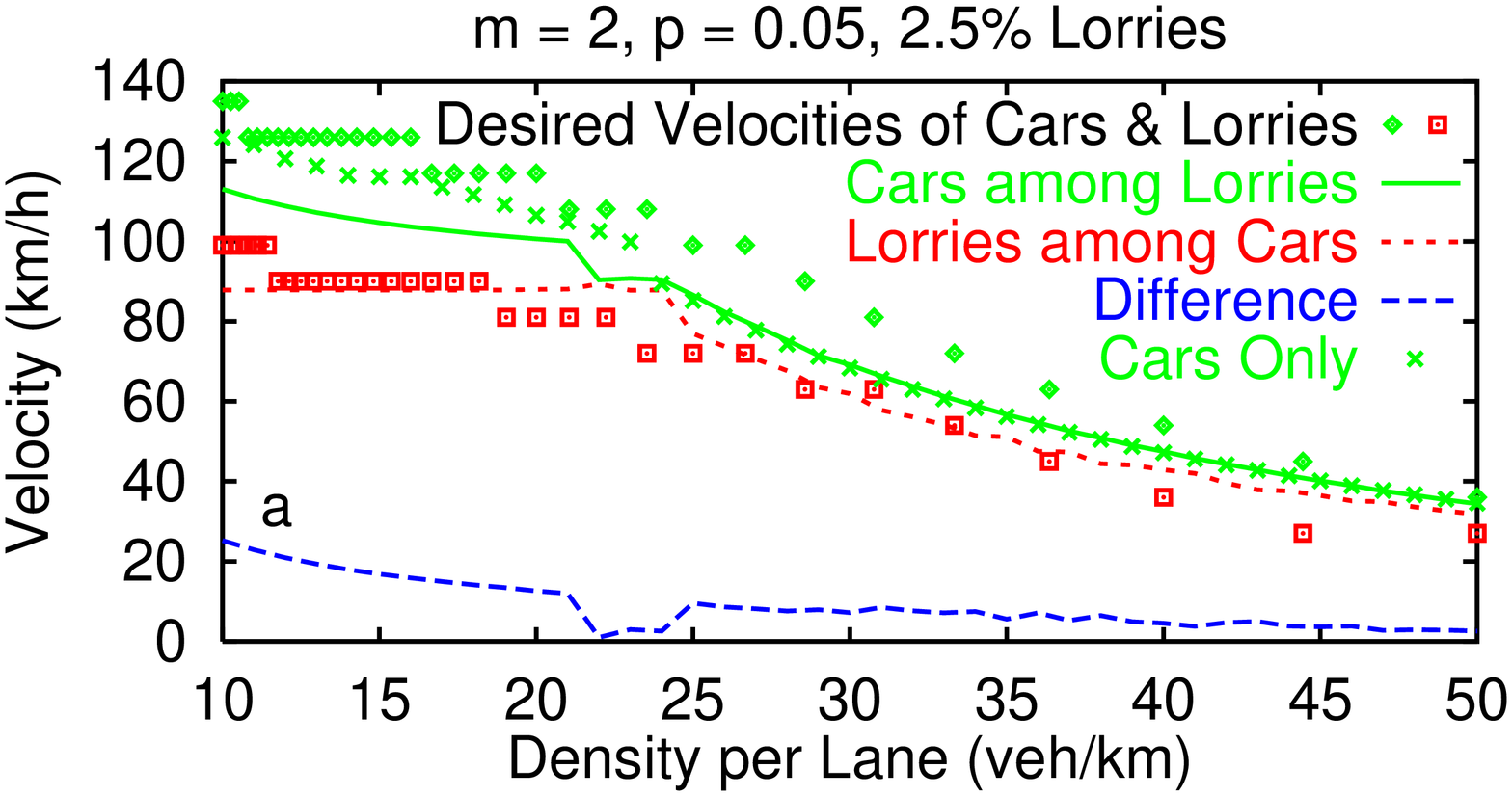}}
\put(1.85,1.81){\epsfig{height=2\unitlength, angle=-90,
      file=fig2_inset.ps}} 
\end{picture}
\hspace*{-0.15\unitlength}
\epsfig{height=4.5\unitlength, angle=-90,
      file=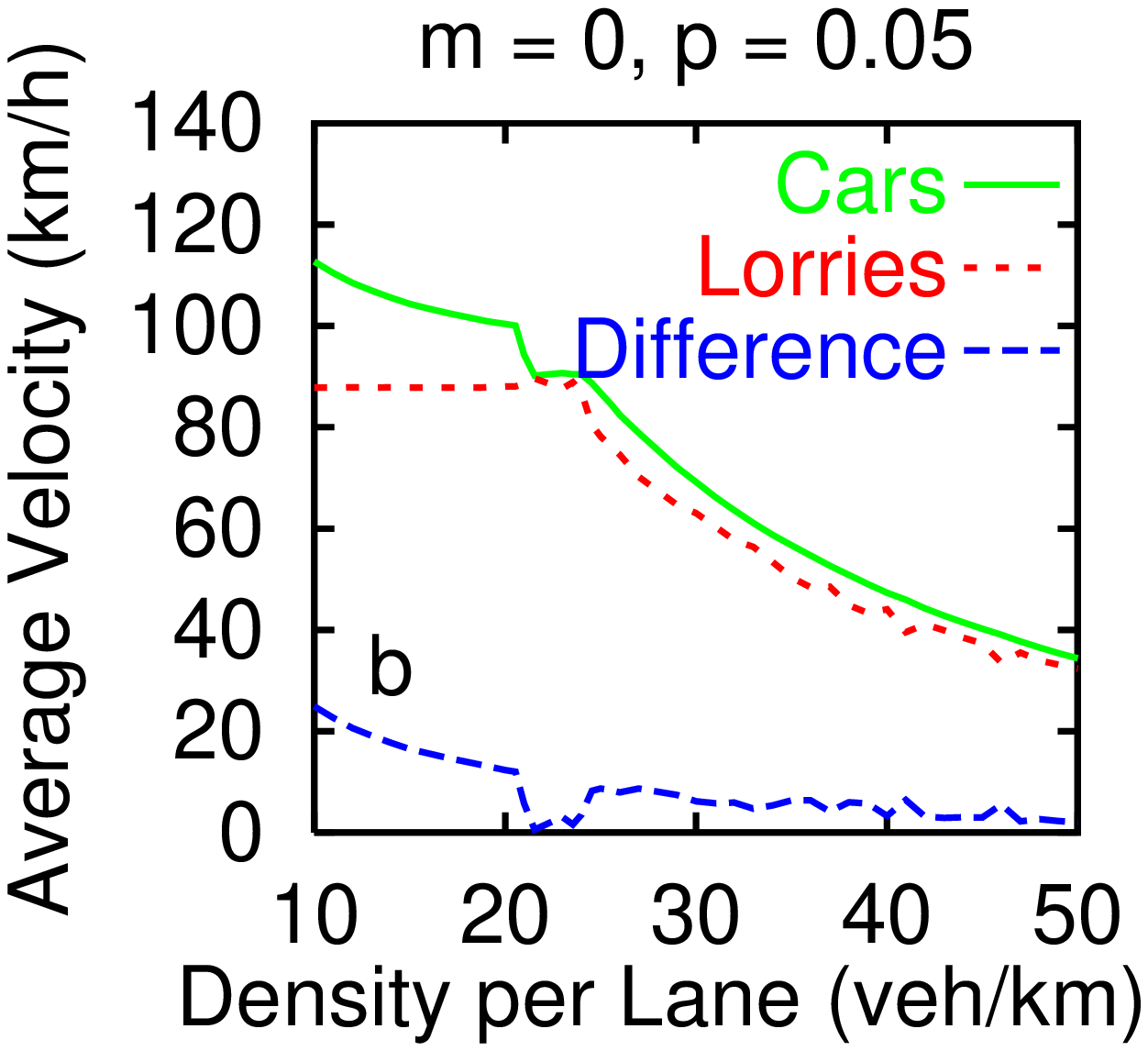} \hspace*{-1.2\unitlength}
\epsfig{height=4.5\unitlength, angle=-90,
      file=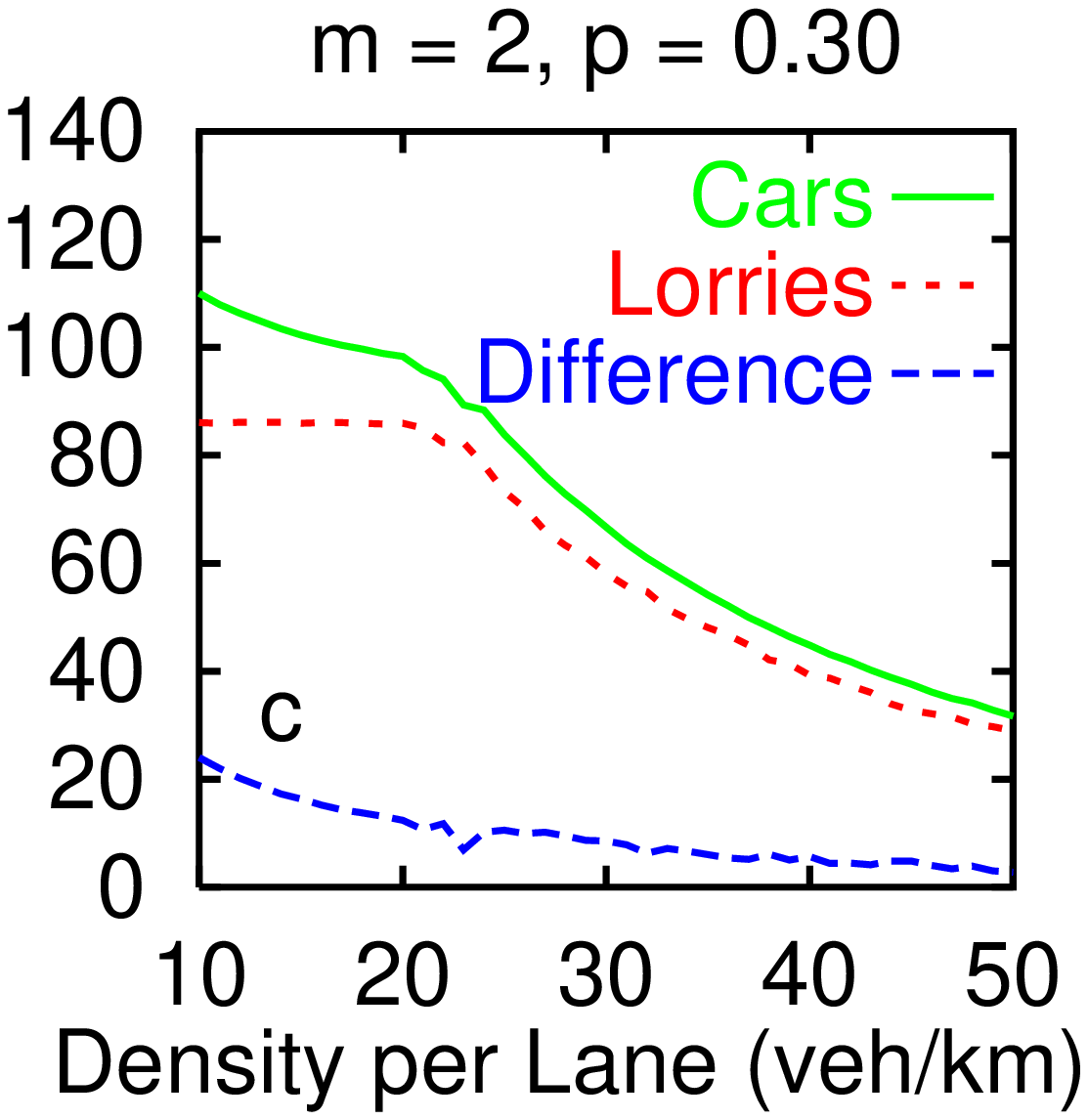}
\end{center}
\caption[]{Numerically determined average velocities for cars 
(---) and lorries (-~-~-) in dependence of the overall vehicle density.
(a) One observes two transitions: 1. Up to a density of
24\,veh/km, the average velocity of lorries remains unchanged.
This is analogous to 
the dynamics of mixed one-lane traffic\,\cite{Krug}: Because of their slower
speed, lorries feel smaller local densities as long as the average velocity
of pure car traffic ($\times$) stays above the maximum velocity of lorries, 
so that the road is not completely used by the vehicular space requirements
at this speed. Then, it falls significantly with growing density
to maintain safe headways. This causes an instability of traffic 
flow\,\cite{Bando,book} resulting in a formation 
of stop-and-go traffic. 2. At a density of
about 21.5\,veh/km, the average velocity of cars
almost drops to the average velocity of lorries, leading to a
distinct minimum in the difference of both curves
(--~--). As illustrated by the inset
in greater density resolution, 
this novel transition seems to be quite sharp. 
Between 21.5 and 24\,veh/km, the vehicles move like a solid block.  
Obviously, this is not enforced by the difference between the
assumed dependencies of the desired velocities 
of cars ($\Diamond$) and lorries ($\Box$) on the local density ahead 
of them, defined by the inverse vehicle distance.\\ 
The parameter values were chosen as in 
Fig.~\ref{F1}, but the observed effects
are not very sensitive to their particular
choice. Different values of $m$ give the same
qualitative results (b). A higher proportion of lorries 
leads to an earlier transition to
the solid-like behaviour. Increasing the fluctuation parameter $p$
causes a smoother, but still visible transition, until it disappears for
$p\approx 0.3$ (c). A similar thing holds for the width of velocity or
parameter distributions characterising cars and lorries. 
\label{F2}}
\end{figure}

In what follows we exhibit a new type of collective behaviour that we
discovered when studying the dynamics of a diverse set of vehicles, such as
cars and lorries, travelling through a two-lane highway with 
different velocities. As the density of
vehicles in the road increases, there is a transition into a highly coherent
state characterised by all vehicles having the same average velocity and a
very small dispersion around its value. 
The transition to this behavior becomes apparent when looking at
the travel time distributions of cars and lorries, 
comprising the overall dynamics on
a freeway stretch (Fig.~\ref{F1}). These were obtained by
running computer simulations using a discretised follow-the-leader 
algorithm\,\cite{FL,Bando}, which 
distinguishes two neighbouring lanes $i$ 
of an unidirectional freeway. Both are subdivided into sites
$z\in \{1,2,\dots,L\}$ 
of equal length $\Delta x = 2.5$\,m. Each
site is either empty or occupied, the latter case representing the back of 
a vehicle of type $a$ (e.g. a car or lorry) 
with velocity $v = u \Delta x / \Delta t$. Here, 
$u \in \{0,1,\dots, u_a^{\rm max} \}$ is the number of sites that the vehicle
moves per update step $\Delta t = 1$\,s.
Cars and lorries are characterised by different `optimal' or `desired' 
(i.e.\ maximally safe) velocities $U_a(d_+)$ with which
the vehicles would like to drive at a distance $d_+$ to the
vehicle in front (see symbols in Fig.~\ref{F2}). Their lengths $l_a$
correspond to the maximum distances satisfying $U_a(l_a) = 0$. At times 
$T\in \{1,2,\dots\}$, i.e. every time step $\Delta t$,
the positions $z(T)$, velocities $u(T)$,
and lanes $i(T)$ of all vehicles are updated in parallel.
We have ruled out synchronisation artefacts\,\cite{upd} by this 
update method, which is appropriate for flow simulations\,\cite{Nag}.
\par
Denoting the position, velocity, and distance
of the respective leading vehicle (+) or following vehicle ($-$)
on lane $i(T)$ by $z_\pm$,
$u_\pm$, and $d_\pm = |z_\pm - z|$, in the adjacent lane 
by $z'_\pm$, $u'_\pm$, and $d'_\pm = |z'_\pm - z|$, the
successive update steps are:
1.~{\em Determine the potential velocities} $u(T+1)$ and $u'(T+1)$ on
the present and the adjacent lane according to the 
{\em acceleration law}\,\cite{zellauto}
\begin{displaymath}
u^{(\prime)}(T+1) = \Big\lfloor \lambda
 U_a\big(d^{(\prime)}_+(T)\big) + (1-\lambda )
 u(T) \Big\rfloor \, ,
\end{displaymath}
where the floor function $\lfloor x \rfloor$ is defined
by the largest integer $l\le x$. This describes the typical
follow-the-leader behaviour of driver-vehicle units. Delayed by the
reaction time $\Delta t$, they tend to move
with their desired velocity $U_a$, but
the adaptation takes a certain time $\tau = \lambda \, \Delta t$ because
of the vehicle's inertia. 
\par
2.~{\em Change lane} in accordance with, for simplicity,
{\em symmetrical} (`American') rules,
if the following incentive and safety criteria\,\cite{zwei} 
are fulfilled: Check if the distance
$d'_-(T)$ to the following vehicle in the neighbouring lane
is greater than the distance $u'_-(T+1)$
that this vehicle is expected to move within the reaction time $\Delta t$
(safety criterion 1). If so, the difference
$D = d'_-(T) - u'_-(T+1) > 0$
defines the backward surplus gap. Next, look if you
could go faster in the adjacent lane (incentive criterion). Finally, make
sure that the relative velocity $[u'_-(T+1) - u'(T+1)]$
would not be larger than $D+m$ (safety criterion~2),
where the magnitude of the parameter $m \ge 0$ is a measure of how
aggressive %($m>0$) of timid ($m<0$) 
drivers are in overtaking (by possibly enforcing deceleration manoeuvres).
\par
3.~If, in the updated lane $i(T+1)$, the corresponding
potential velocity $u(T+1)$ or $u'(T+1)$ is positive,
diminish it by 1 with probability $p$ in order to 
account for delayed adaptation and the variation of
vehicle velocities. 
\par
4.~Update the vehicle position according to the {\em equation of motion}
$z(T+1) = z(T) + u(T+1)$.
\par
Despite its simplifications, this model is in good agreement with
the empirical known features of traffic flows, 
and it can be well calibrated\,\cite{zellauto}: 
$\lambda$ and $U_a(d_+)$ determine 
the approximate velocity-density relation and the instability region. 
The typical outflow from traffic jams and their characteristic dissolution
velocity\,\cite{Kerner1} can be enforced by  
$\Delta t$ and $\Delta x$. 
The average distance between successive traffic jams  
increases with smaller $p$. $m$ allows to calibrate
the lane-changing rates. %\,\cite{Leutz}.
Our simulations started with uniform distances among the vehicles
and their associated desired velocities.
The lorries were randomly
selected. Since our evaluations started after a transient period of
one hour and extended over another four hours, the results are largely
independent of the initial conditions. 
\par
\begin{figure}
\begin{center} \hspace*{0.1\unitlength}
\epsfig{height=7.2\unitlength, angle=-90,
      file=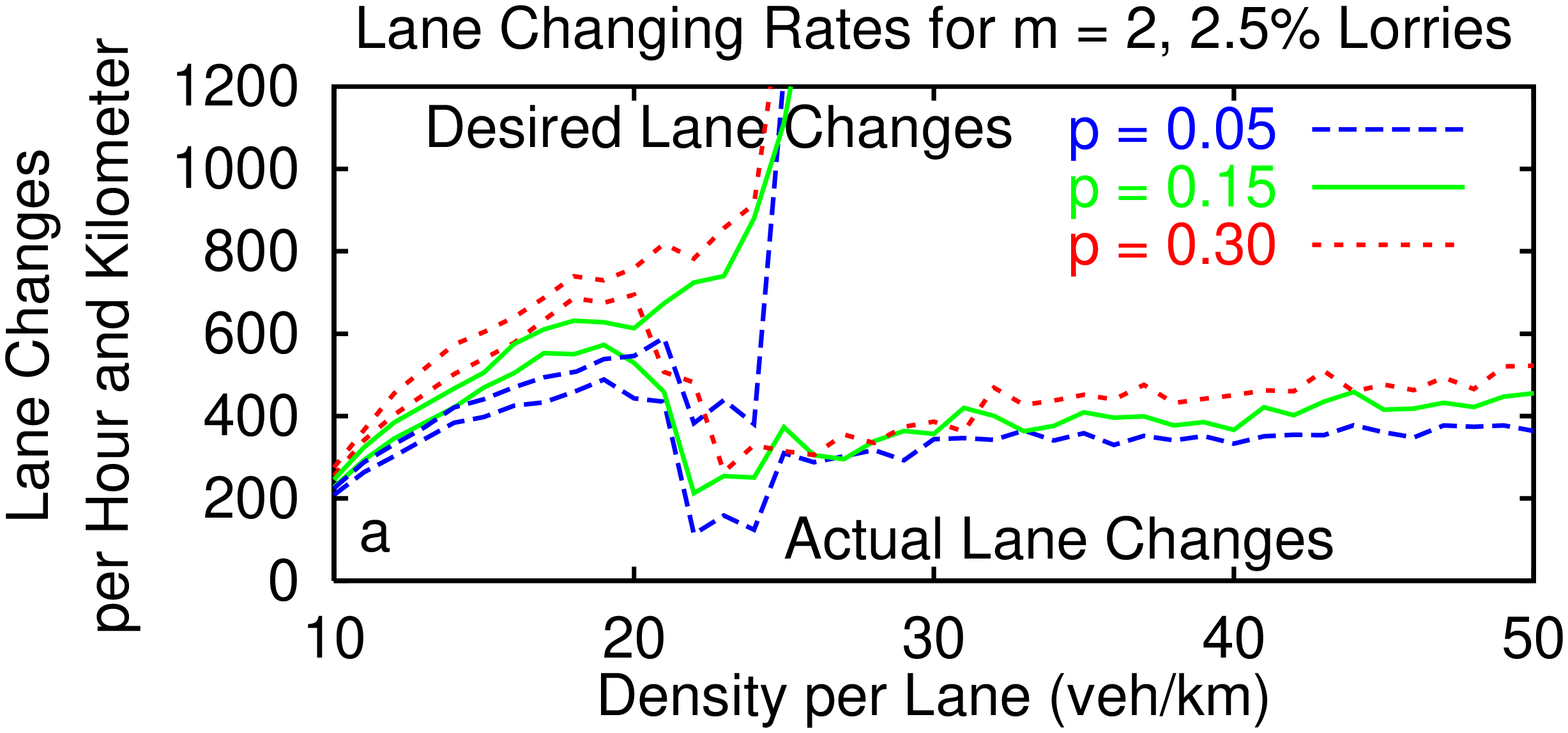} \\
\hspace*{-0.4\unitlength}
\epsfig{height=4\unitlength, angle=-90,
      file=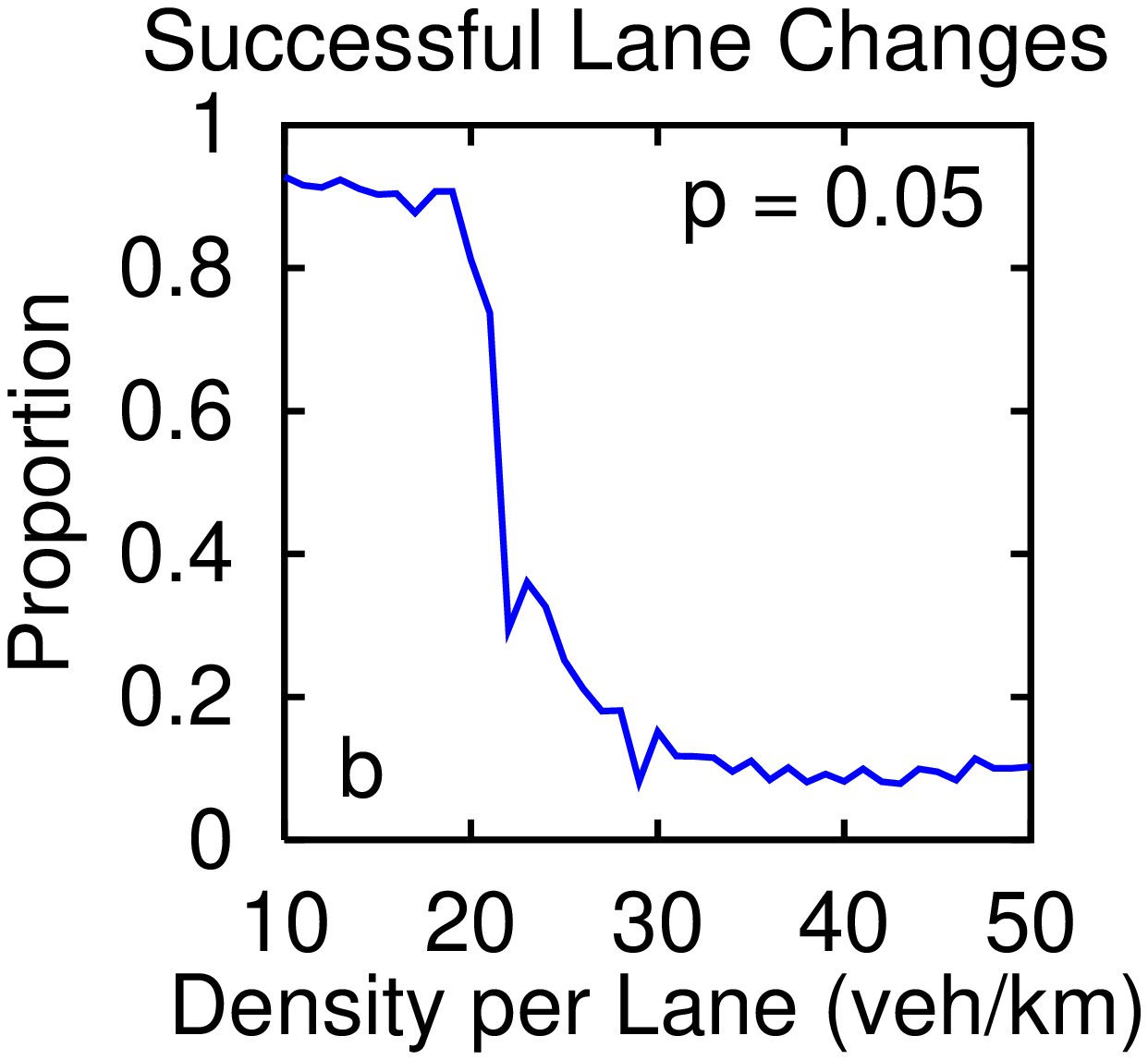} \hspace*{-0.6\unitlength}
\epsfig{height=4\unitlength, angle=-90,
      file=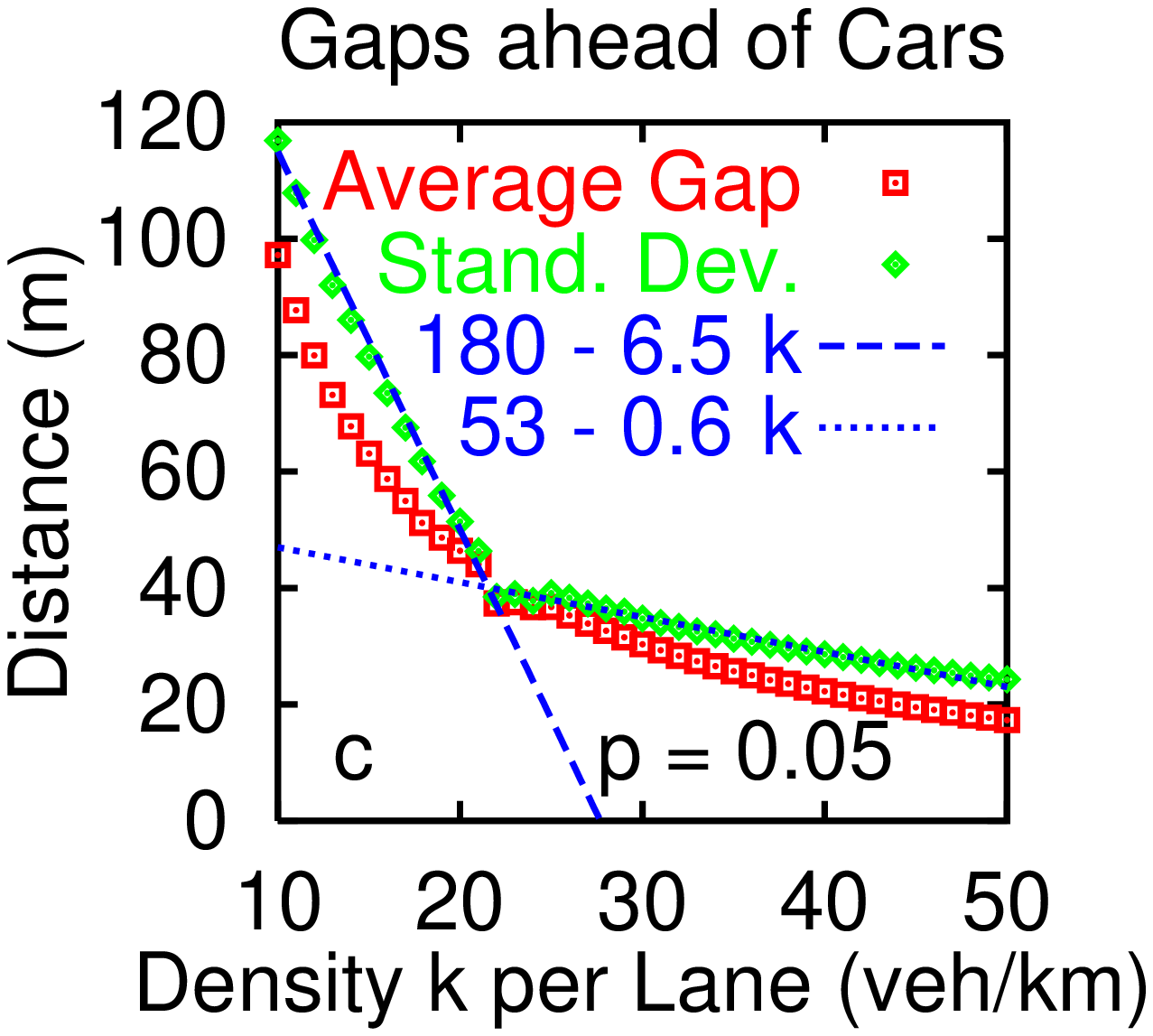}
\end{center}
\caption[]{(a) The simulated {\em actual} lane-changing rates 
break down in the density range of coherent motion.
Here, the parameters are the same as in Figs.~\ref{F1} and \ref{F2}.
For stronger fluctuations $p$, the minimum is less pronounced, but 
still noticeable up to $p < 0.3$. Smaller values of $m$ reduce 
the number of lane changes, but do not prevent the solid-like state.
We have checked whether the lane-changing rate breaks down
because, with identical velocities of
cars and lorries in both lanes, there may be no advantage of lane changing.
However, if the rates of {\em desired} lane changes (according to 
the incentive criterion) are reduced at all, they still keep a high level.
(b) The transition point around 21.5\,veh/km is characterised by a
rapid decay of the proportion of successful
lane changes (i.e. the quotient between actual and desired lane changes). 
(c) With growing density, not only the average of the 
gaps in front of cars decreases 
($\Box$), but also their standard deviation ($\Diamond$). 
Opportunities for lane-changing are rapidly diminished when gaps 
of about twice the safe headway required for lane-changing cease to
exist. This relates to a significantly smaller slope of the 
gaps' standard deviation after the solid-block transition (broken lines).\\ 
The breakdown of the lane-changing rate seems to imply a decoupling of the
lanes, i.e. an effective one-lane behavior. However, this 
is already the {\em result} of a self-organisation process based on
two-lane interactions, since
any significant perturbation of the solid block state (like different 
densities in the neighbouring lanes or velocity variations) will cause frequent
lane changes. By filling large gaps, the gap distribution is considerably
modified (also compared to mixed one-lane traffic\,\cite{Krug}).
This will eventually reduce possibilities for lane changes, 
so that the solid-like state is restored. 
\label{F3}}
\end{figure}

Investigating the density-dependent average velocities of cars and lorries
yields further insight into the solid-like state (Fig.~\ref{F2}).
For small $p$ one finds that, at a certain `critical' density, 
the average speed of cars decays significantly 
towards the speed of the lorries, which is still close to their maximum 
velocity. At this density, the freeway space is almost used up by 
the safe vehicle headways, so that sufficiently large gaps for 
lane-changing can only occur for strongly varying vehicle velocities
(like for large $p$). However, since the speeds of cars and lorries 
are almost identical in the solid-like state, the lane-changing rate 
drops by almost one order of magnitude (Fig.~\ref{F3}). Consequently, without
opportunities for overtaking, all vehicles have to move
coherently at the speed of the lorries, which closes the feedback-loop 
that causes the transition. 
%\par
The solid-like flow does not change by
adding vehicles until the whole freeway is saturated by the vehicular
space requirements at the speed of the lorries. Then, the vehicle
speeds decay significantly to maintain safe headways. The onset of
stop-and-go traffic at this density produces largely varying gaps,
so that overtaking is again possible and the coherent state is
destroyed. For large $p$, we do not have a breakdown of the lane-changing 
rate at the critical density and, hence, no coherent state. Nevertheless,
lane-changing cars begin to interfer with the lorries, so that the average
velocity of lorries starts to decrease with growing density 
{\em before} the average car velocity comes particularly close to 
it (Fig.~\ref{F2}c).
\par
\begin{figure}
\begin{center}
\hspace*{-0.2\unitlength}
\epsfig{width=7\unitlength, angle=-90,
      file=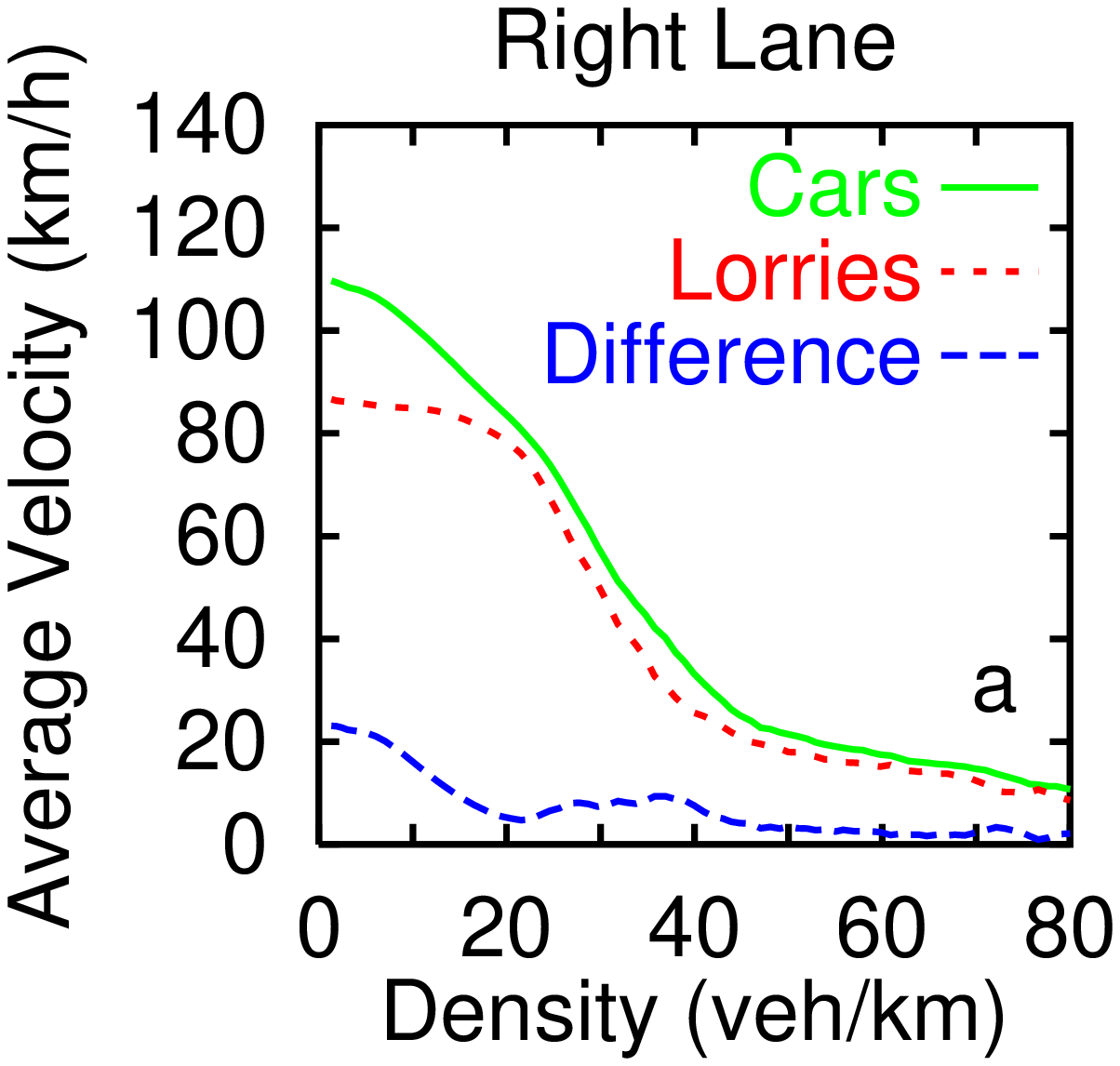} \hspace*{-0.9\unitlength}
\epsfig{width=7\unitlength, angle=-90,
      file=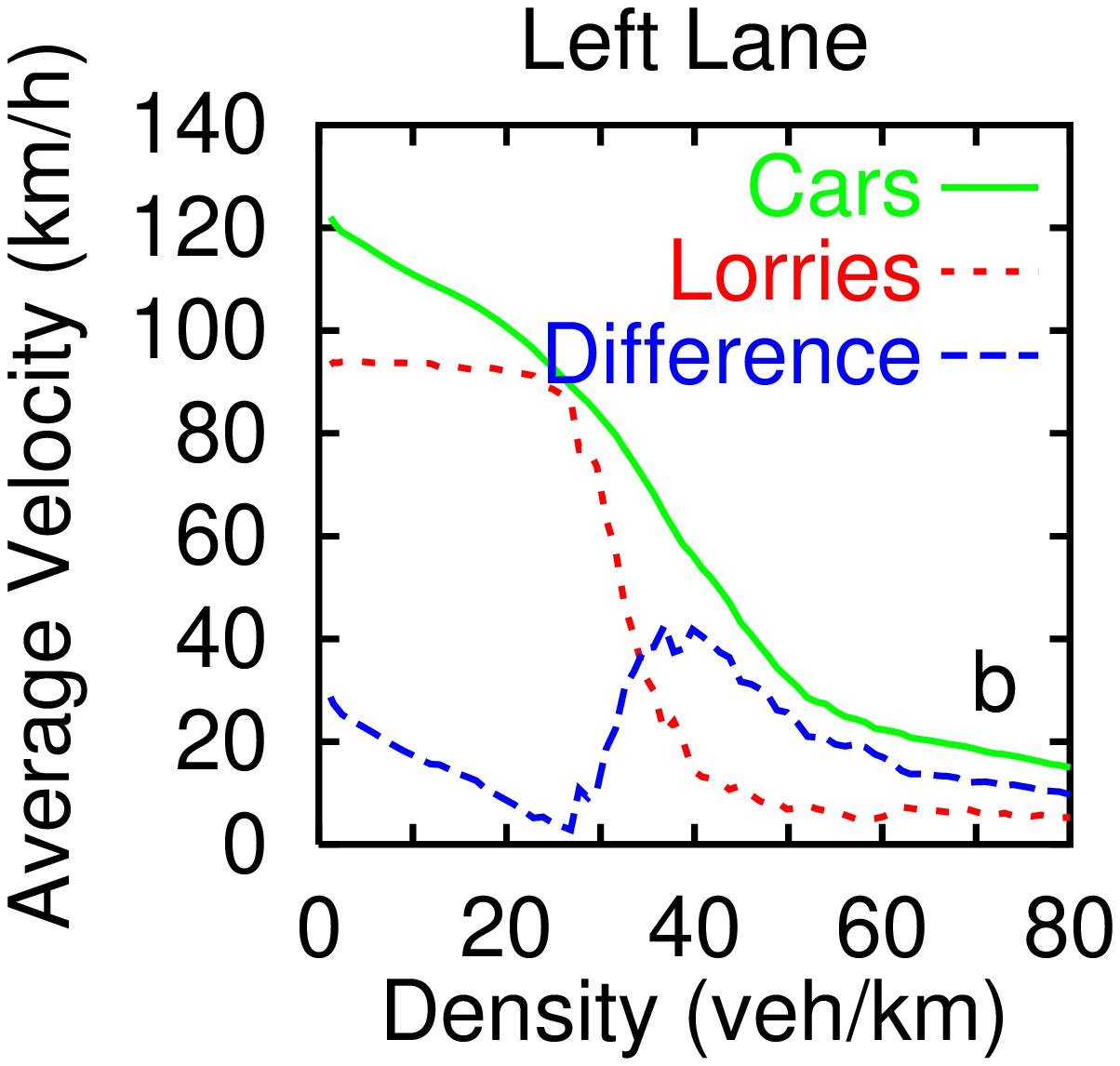} \\ \hspace*{-0.4\unitlength}
\epsfig{width=7\unitlength, angle=-90,
      file=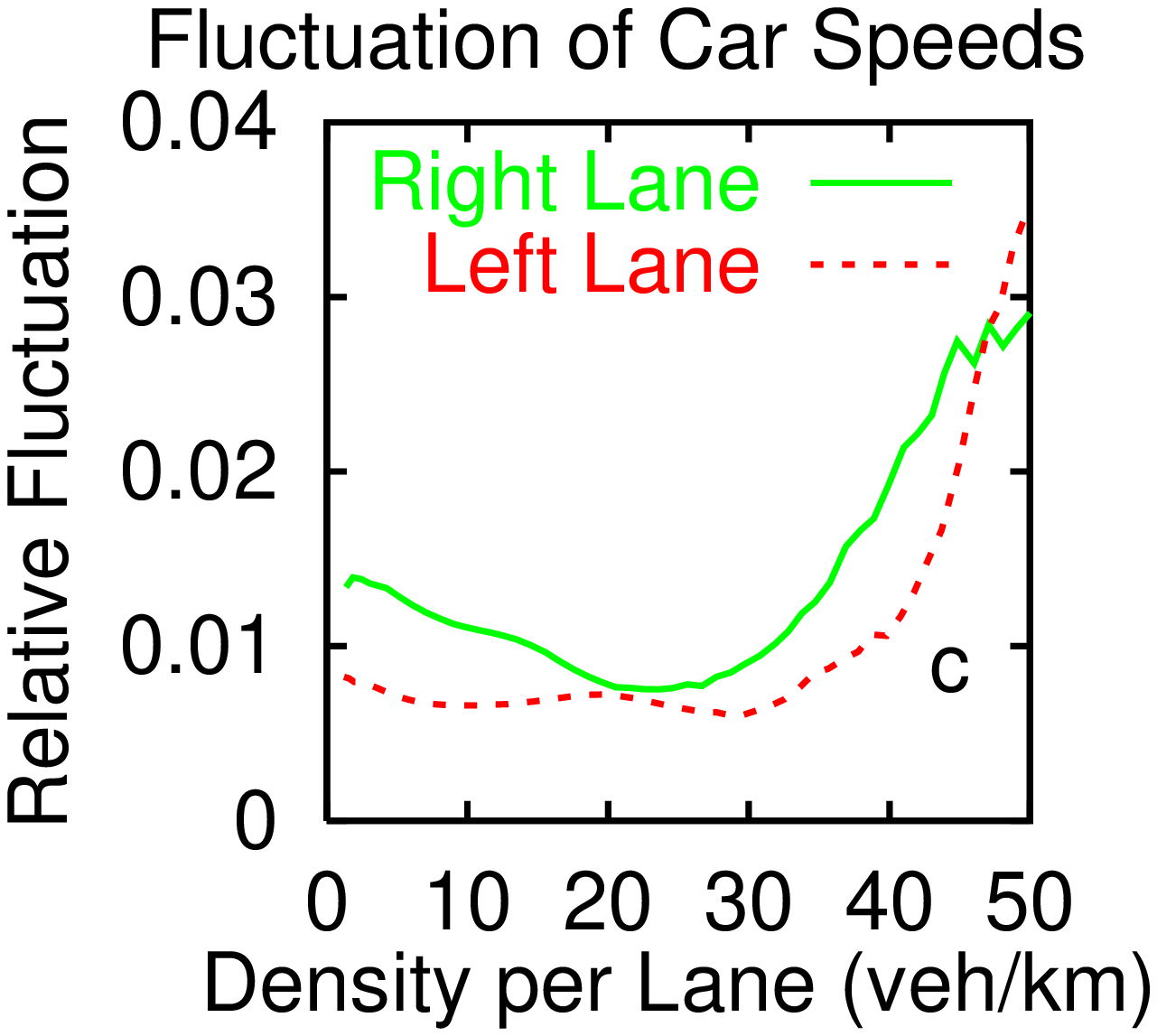} \hspace*{-0.4\unitlength}
\epsfig{width=7\unitlength, angle=-90,
      file=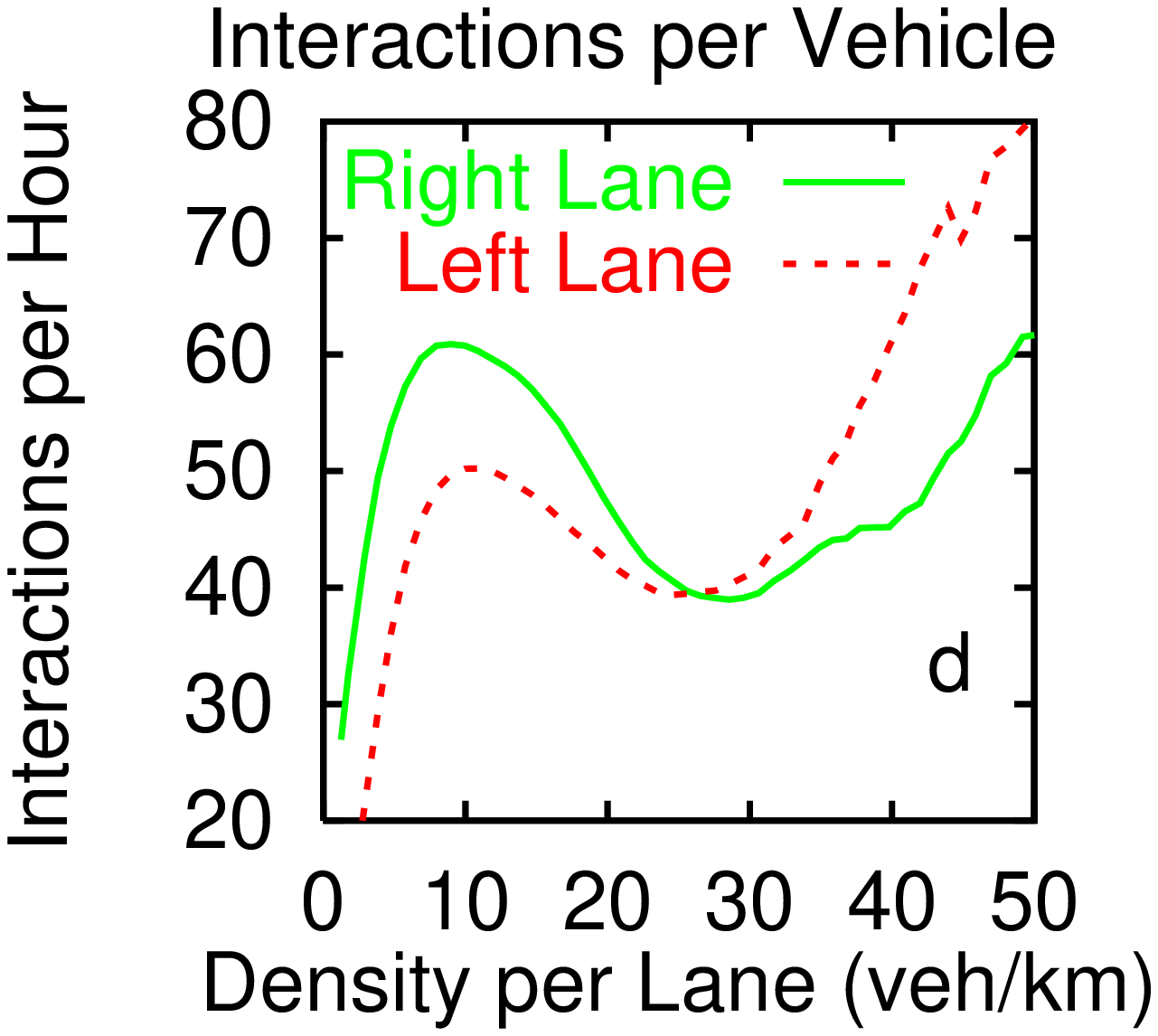}
\end{center}
\caption[]{Mean values of one-minute averages that were determined from 
single-vehicle data of mixed traffic on the Dutch two-lane 
freeway A9 on 14 subsequent days. Illustrations (a) and (b) depict 
the density-dependent average velocities
of cars (---) and lorries (-~-~-) in the right and left lane, respectively.
Their differences (--~--) show the predicted minimum 
at densities of about 25\,veh/km, up to which the velocity of lorries 
is almost constant. 
(c) The relative fluctuations of
car speeds (defined as velocity variance divided by the square of
average velocity) display minima at the same densities, which 
points to a more coherent motion in the car fraction.
(d) The interaction rates per vehicle
show a minimum around 25\,veh/km as well. This corresponds to 
a decreased {\em relative} velocity among successive vehicles. 
The interation rate is
defined by the average of $\min(\Delta v/d_+,0)$, where
$\Delta v$ denotes the velocity difference to the vehicle in front.\\
We point out that the above data support the predictions of our model 
despite its simplifications. 
In particular, this concerns the asymmetry of European lane-changing rules,
which imply that overtaking is only allowed
in the left lane, 
but vehicles should switch back to the right lane as soon as
possible. At least at velocities of about 80\,km/h or below, vehicles
also pass in the right lane with small relative velocities to vehicles in
the left lane. Nevertheless, the rate of lane changes from and
to each lane is, on average, the same. 
One result of the mentioned asymmetry, however, is a smaller fraction of 
lorries in the left lane, which diminishes the related minimum in 
(c). Another consequence is the tendency of having a
higher average  car velocity in the left lane, which pronounces the features
in (b) compared to (a).
\label{F4}}
\end{figure}

As shown in  Figure~\ref{F4}, our prediction of the transition into
a coherent state is
supported by empirical data obtained from highway traffic in the Netherlands.
Clearly, the difference between the average velocities
of cars and lorries shows the predicted minimum
at a density around 25 vehicles per kilometer and lane, where the
average car velocity approaches the constant velocity 
of the lorries (Fig.~\ref{F4}a,b). The fact that the empirically observed
minimum is less distinct than in Fig.~\ref{F2}a can be reproduced 
by higher values of the fluctuation parameter ($p \approx 0.15$) and
points to a {\em noisy} transition.
This interpretation is also supported by the
relative fluctuation of vehicle speeds (Fig.~\ref{F4}c), which shows a
minimum at the same density, while remaining finite. A similar result 
is obtained for the interation rates of vehicles (Fig.~\ref{F4}d).
\par
In summary, we have presented a novel effect in highway traffic that
consists in the formation of coherent motion out of a disorganised 
vehicle flow by competitive interactions. 
The predicted solid-like state is supported 
by real highway data, and our interpretation 
of the effect suggests that it is largely
independent of the chosen driver-vehicle model
(although the transition may be less sharp in a continuous model).
It would be interesting to see if the spontaneous
appearance of coherent states is also found in other social or
biological systems, such as pedestrian crowds\,\cite{Fuss}, 
cell colonies\,\cite{Jacob},
or animal swarms\,\cite{Vicsek}. 
\par
We conclude by noting that the coherent state of vehicle motion considerably reduces the main
sources of highway accidents: differences in vehicle speeds and 
lane-changes\,\cite{Lave}. It is
also associated with maximum throughput in the highway 
and located just before the
transition to unstable traffic flow. Thus,
at a practical level it is desirable to implement traffic rules and 
design highway controls that will lead to traffic moving like a
solid block. Close to the transition point the formation of this coherent state %of motion 
could be supported by traffic-dependent lane-changing restrictions and 
variable speed limits or by automatic vehicle control systems.
Compared to American (symmetric)
lane-changing rules, European ones 
seem to be less efficient: An asymmetric lane usage, 
where lorries mainly keep on one lane and overtaking is carried out on 
the other lane(s), motivates car drivers 
to avoid the lorry lane, so that the effective 
freeway capacity is reduced up to 25\%. 

{\em Acknowledgments:}
D.H. wants to thank the DFG for financial support
by a Heisenberg scholarship. He is also grateful to Henk Taale and
the Dutch {\it Ministry of Transport, Public Works and Water Management}
for supplying the freeway data, which were evaluated by Vladimir Shvetsov.
\clearpage
%%%%%%%%%%%%%%%%%%%%%%%%%%%%%%%%%%%%%%%%%%%%%%%%%%%%%%%%%%%%%%%%%%%%%%%%%%%%%%
\unitlength8mm
\clearpage
\unitlength14mm
\clearpage
\unitlength16mm
\clearpage
\unitlength8mm
\end{document}